\documentclass[conference]{IEEEtran}
\usepackage{blindtext, graphicx}
\usepackage{chemmacros}

\ifCLASSINFOpdf
\else
\fi
\usepackage{amssymb}

\usepackage{listings}
\usepackage{color}

\definecolor{dkgreen}{rgb}{0,0.6,0}
\definecolor{gray}{rgb}{0.5,0.5,0.5}
\definecolor{mauve}{rgb}{0.58,0,0.82}

\usepackage{fancyhdr}

\usepackage{hyperref}

\begin{document}
\title{DNA-based chemical compiler}
\author{\IEEEauthorblockN{Shalin Shah\IEEEauthorrefmark{1}\IEEEauthorrefmark{2} and Manish Gupta\IEEEauthorrefmark{2}}
\IEEEauthorblockA{\IEEEauthorrefmark{1}Department of Electrical \& Computer Engineering, Duke University, Durham, US - 27701\\
\IEEEauthorrefmark{2}Dhirubhai Ambani Institute of Information \& Communication Technology, Gandhinagar, India - 382007  \\
Email: mankg@computer.org}}

\maketitle

\begin{abstract}
Marcello, in 1997, formally proved that chemical kinetics can make a universal computer i.e they can replicate any digital circuit~\cite{magnasco1997chemical}. Recently, Soloveichik \textit{et al.} showed that chemical kinetics can perform fast and reliable Turing Universal computations~\cite{soloveichik2008computation}. To simulate the behavior of chemical reactions, Sean \textit{et. al.} developed a software called CAIN which represents chemical reactions in XML format~\cite{seanCAIN}. In this work, we have made an attempt to create trans-compiler which can take python like code as input and gives CAIN supported chemical reactions file as output. This can be compared to generating assembly code from a high level programming language. Additionally, Soloveichik \textit{et al.} also showed DNA as a universal primer for implementing CRN's~\cite{soloveichik2010dna} and Andrews Phillips \textit{et. al.} developed Visual DSD programming language for simulating all the possible DSD reactions~\cite{phillips2009programming}. CRN2DSD, a software developed by Manish Gupta's team, can, already, convert a CAIN file to Microsoft's Visual DSD code i.e assembly level to machine level. Hence, our attempt to convert high level code to assembly code takes us one step closer to completing the dream of making a chemical compiler.
\end{abstract}


\IEEEpeerreviewmaketitle

\section{Introduction}
DNA has found variety of applications such as archival storage~\cite{shah2013dnacloud}, drawing on molecular canvas~\cite{wei2012complex}, large scale logic circuits~\cite{qian2011scaling}, complex computing problems~\cite{DBLP:journals/corr/ShahDG15} and making structures at nano-scale with help of DNA origami~\cite{rothemund2005design}. This is largely because of its programmable behavior and easy synthesis~\cite{qian2011scaling}. In 2011, Qian \textit{et. al.} showed that efficient DNA strand displacement - a proposed technique for computation using DNA - phenomenon is Turing Universal~\cite{qian2011efficient}. Later, a motley of circuits such as join/ fork, transducer, square-root, AND, OR, NOT etc. have been implemented using the DNA strand displacement (DSD) phenomenon~\cite{qian2011scaling, cardelli2013two, jiang2013digital}. Andrew Phillips \textit{et. al.} also developed a programming language called Visual DSD to simulate the behavior of DSD~\cite{phillips2009programming}. 

Soloveichik \textit{et al.} showed that chemical kinetics can perform fast and reliable Turing Universal computations~\cite{soloveichik2008computation} and, additionally, he also showed that DNA can form a universal substrate for chemical kinetics. To simulate chemical reactions, Sean \textit{et. al.} developed a software called CAIN which represents chemical reactions in XML format~\cite{seanCAIN}. Therefore, Manish Gupta and his group developed CRN2DSD which can convert chemical reactions file (CAIN XML file) to its corresponding DNA based Visual DSD file. This is equivalent to converting assembly code (chemical reactions) to machine code (reversible DSD reactions). However, the problem is to create chemical reaction for any given computation. Therefore, we have made an attempt to write a trans-compiler which can take python like file as input and generate its CAIN XML file containing chemical reactions. This is similar to converting high level language to assembly language. By successfully implementing this step, and combining it with CRN2DSD, one can realize the dream of chemical compiler. 

\begin{figure}[ht!]
\centering
\includegraphics[scale=0.4]{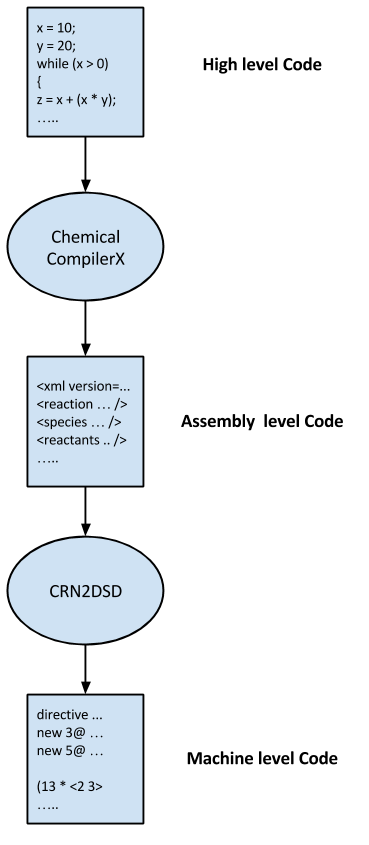}
\centering
\caption{The abstract level representation of a chemical compiler.}
\label{imageCSummary}
\end{figure}

Attempts have already been made in this direction by creating stochastic rate-independent chemical reactions for increment/decrement, copying, clearing, doubling etc.~\cite{senum2011rate}. This was the first attempt in the direction of \textit{synthesis}: given a task, how can one write chemical reactions for the same? Earlier attempts such as~\cite{qian2011efficient} were made in the direction of \textit{analysis}: Can we classify the reactions according to its behavior? Other attempts include Verilog based compiler named Verilog Elements for Register-Based Biochemistry (VERB) which can convert a Verilog file to corresponding set of stochastic chemical reactions on the basis of clocking mechanism. Fig~\ref{imageCSummary} shows the abstract form of our idea.

In this report, we give brief idea about our compiler and its working. In Section 2, we give a brief idea as to what stochastic model for CRN is. In Section 3, we show the workflow of the compiler along with intermediate outputs. In Section 4, we mention the basic rate independent chemical reactions required. In Section 5, there are important points about compiler which one should know. In Section 6, intermediate output graph i.e AST is shown. Section 7 shows how final output will look while section 8 shows how compiler can do error handling. In Section 9, we mention the problems which we are not able to figure out yet. 

\section{Background}
According to~\cite{gillespie1977exact}, chemical kinetics can be classified into following models: 

\begin{enumerate}
\item Deterministic Model
\item Stochastic Model
\end{enumerate}

We are interested in discrete, stochastic model, mainly, because it represents a limited number of reactants and products unlike deterministic model. The deterministic models works well in cases where there are a large number of molecules of reactants and products, like in macroscopic reactions, which vary continuously over time~\cite{soloveichik2010dna}. Ordinary differential equations are used to represent their concentration dynamically. Stochastic model, on the other hand, keeps track of exact number of molecules like in cell reactions~\cite{soloveichik2010dna}. Note that when molecular counts are large stochastic model changes to deterministic model~\cite{soloveichik2008computation}. To substantiate the small count of molecules in cellular reaction, one can consider E.coli bacteria whose 80\% genes have less than a hundred copies per cell~\cite{guptasarma1995does}.

Consider a solution containing n reactions and m species. Every reaction has non-negative integer molecules of reactants, represented by a vector $\textbf{r} = (r_1, r_2, \ldots, r_x)$, and products, represented by a vector $\textbf{p} = (p_1, p_2, \ldots, p_y)$, where $\textbf{r} \epsilon \mathbb{N}^x, \textbf{p} \epsilon \mathbb{N}^y$ and $\mathbb{N} = \{0, 1, 2, 3, ..\}$. The rate of reaction is proportional to concentration of reactants and rate constant. For example, consider the reaction: 

\ch{start + start + var ->[ k ] var + start} \hfill (1)

In this reaction, we consume 2 molecules of start and one molecule of var to produce one molecule of var and start at rate k. Not that var is used as a catalyst in this reaction i.e it is consumed and produced back so its quantity remains same. In this work, we follow \cite{senum2011rate} for rate of reactions so there are only two rates \textit{fast} and \textit{slow}, where \textit{slow} is relatively slower than \textit{fast}. Say for example, we have four reactions, two of them have rate constants \textit{fast} and other two having rate constants \textit{slow}. It doesn't matter how \textit{fast} are the fast rates. Additionally, both of them need not necessarily be the same all we need is that they should be faster than \textit{slow} rate. Since all reactions are implicitly reversible, an important assumption here is that reverse rates for all the reactions are much slower than forward rate.

\section{Workflow of Compiler}
\begin{figure}[ht!]
\centering
\includegraphics[scale=0.45]{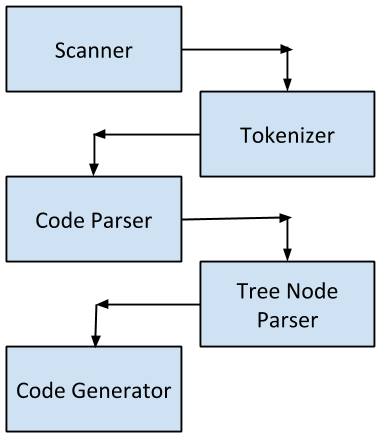}
\centering
\caption{The work-flow of chemical compiler.}
\label{imageCWorkflow}
\end{figure}

This compiler, like traditional compiler, follows steps shown in Fig.~\ref{imageCWorkflow} to convert input code from high level language to assembly language.

\begin{enumerate}
\item Take the python like source code as input and divide the source code in to tokens. Tokenization is handled by Flex.
\item After the code is tokenized, each line parsed according to the context free grammar (CFG) of the language. The grammar rules were defined in Bison.
\item The output of parser is a tree which is called Abstract Syntax Tree (AST). This tree is parsed node by node.
\item Generate chemical reactions in XML format for every node traversed.
\end{enumerate}

In traditional compilers, there is an additional step of optimization which we haven't covered in mine since it is not required as of now.

\begin{figure*}
\centering
\includegraphics[width=0.8\textwidth]{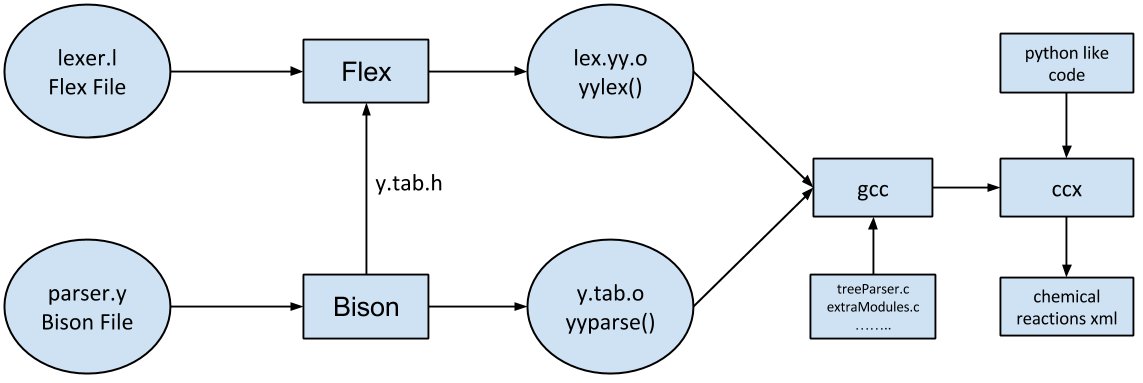}
\centering
\caption{The work-flow of building process of compiler. Output of bison is a header file and an object file which combines with Lexer to give another object file. Both these object files and other files used in compiler are then combined to make an executable file ccx.}
\label{imageWorkflow}
\end{figure*}

\section{Chemical Reactions}
In this compiler, we have tried to implement very basic operations such as addition, subtraction and multiplication. These operations are implemented using some fundamental reactions. For example the simplest operation in any compiler is assignment of variable which is implemented using copy reactions. Therefore, in this section, we mention the fundamental reactions required to implement operations such as addition and multiplication. We have used the stochastic Chemical Reaction Network (sCRN) model since we need specific non-negative integer amounts of molecules. Additionally, this kind of behavior is also exhibited by cell reactions. To know more about sCRN, refer~\cite{cook2009programmability}.

\subsection{Clear Reactions}
This is most basic reaction used in compiler to set value of the variable zero i.e $x = 0$.

\ch{start + var + done_{ab} ->[ slow ] var' + start} \hfill (1.1)

\ch{start + var_{ab} ->[ slow ] done} \hfill	(1.2)

\ch{done + done ->[ fast ] done} \hfill (1.3)

\ch{done + var' ->[ slow ] done} \hfill (1.4)

\ch{start + done ->[ slow ] done} \hfill (1.5)

These reactions work as follows: In reaction 1, we assume that clear process is not complete so $done_{ab}$ is present. This along with start signal and variable to be cleared form var' species. Once all the variable species are converted, we produce done signal in Reaction 2 by combining start and $var_{ab}$. Reaction 3 keeps a check on quantity of done. Finally, in Reaction 4, we combine done are var' to destroy var'. Reaction 5 converts all the start signal species to done. For detailed explanation on reactions, refer~\cite{senum2011rate}.

\subsection{Copy Reactions}
Assignment operator, undoubtedly, is most commonly used operator in programming languages. The operation $y = x$ can be done using these reactions.

\ch{start + var + done_{ab} ->[ slow ] var' + start} \hfill (2.1)

\ch{start + var_{ab} ->[ slow ] done} \hfill	(2.2)

\ch{done + done ->[ fast ] done} \hfill (2.3)

\ch{done + var' ->[ slow ] done + var + output} \hfill (2.4)

\ch{start + done ->[ slow ] done} \hfill (2.5)

These reactions work as follows: Reactions 1 - 3 and 5 work exactly in the same fashion as clear reactions. The only difference here is Reaction 4 which, instead of destroying original variable, makes output molecules equal to that of var molecules. This is equivalent to operation $x \Rightarrow x + y$. For detailed explanation on reactions, refer~\cite{senum2011rate}.

\subsection{Increment/Decrement Reactions}
These operations is used widely in loops. It is, however, a special case of  $x = y + z$ when $z = 1$. 

\ch{var + start + done_{ab} ->[ slow ] var' + start} \hfill (3.1)

\ch{start + var_{ab} ->[ slow ] done} \hfill	(3.2)

\ch{done + 2 var' ->[ slow ] done + var + var' + var^{rx}} \hfill (3.3)

\ch{var^{rx} ->[ slow ] $\phi$} \hfill (3.4)

\ch{done +  var^{rx}_{ab} + var' ->[ slow ] done} \hfill (3.5.1)

\ch{done +  var^{rx}_{ab} + var' ->[ slow ] done + var +  var} \hfill (3.5.2)

\ch{done + done ->[ fast ] done} \hfill (3.6)

\ch{start + done ->[ slow ] done} \hfill (3.7)

Reaction 3.5.1 is used for decrement while 3.5.2 is used for increment rest of the reactions remain same in both the cases. Except for 3.3, 3.4 and 3.5 reactions, others are similar to copy and clear. In Reaction 3.3, we generate $var^{rx}$ since the absence of this molecule is used to indicate the on-going transfer from var' to var. Once this transfer is completed, the remaining $var^{rx}$ molecules are destroyed with help of Reaction 3.4. For detailed explanation on reactions, refer~\cite{senum2011rate}.

\subsection{Comparison Reactions}
This operation is used in every branching and looping statement in programming language. Therefore, this operation holds a great importance. To start comparison operation, we create copy of our variables since this is a destructive reaction. Thus, consider that using above mentioned copy reactions, we created copy of $var_1$ and $var_2$ as $var_1$' and $var_2$'.

\ch{var1' + var2' ->[ slow ] $\phi$} \hfill	(4.1)

\ch{fuel + var1'_{ab} + var2'_{ab} ->[ slow ] var1'_{ab} + var2'_{ab} + t1} \hfill	(4.2.1)

\ch{fuel + var1' + var2'_{ab} ->[ slow ] var1' + var2'_{ab} + t2} \hfill	(4.2.2)

\ch{fuel + var1'_{ab} + var2' ->[ slow ] var1'_{ab} + var2' + t3} \hfill	(4.2.3)

\ch{var1'_{ab} + var2'_{ab} + t2 ->[ slow ] var1'_{ab} + var2'_{ab}} \hfill	(4.3.1.1)

\ch{var1'_{ab} + var2'_{ab} + t3 ->[ slow ] var1'_{ab} + var2'_{ab}} \hfill	(4.3.1.2)

\ch{var1' + var2'_{ab} + t1 ->[ slow ] var1'_{ab} + var2'_{ab}} \hfill	(4.3.2.1)

\ch{var1' + var2'_{ab} + t3 ->[ slow ] var1'_{ab} + var2'_{ab}} \hfill	(4.3.2.2)

\ch{var1'_{ab} + var2' + t1 ->[ slow ] var1'_{ab} + var2'} \hfill	(4.3.3.1)

\ch{var1'_{ab} + var2' + t2 ->[ slow ] var1'_{ab} + var2'} \hfill	(4.3.3.2)

The reactions work as follows: Reaction 4.1 leaves either copy of varc1 or varc2 or none depending on the relationship between them ($>$, $<$, or $=$). Say for example, if $var1 > var2$, then Reaction 4.2.2 will be fired to creation of $t_2$ and Reaction 4.3.1.1, 4.3.1.2, 4.3.2.2 and 4.3.3.2 will be fired for destruction of unwanted species of $t_1$ and $t_3$. These are generalized reaction and to see how to use these reaction for different comparisons, please refer Table 1 in ~\cite{senum2011rate}. Also, for detailed explanation on reactions, refer~\cite{senum2011rate}.

\subsection{Subtraction Reaction}
In order to subtract a number, we need copy reaction. This operation is implemented by copying the value to be subtracted to a temporary variable and then performing the Reaction 5.1. 

\ch{var + temp ->[ fast ] $\phi$} \hfill	(5.1)

Say for example, if we wish to subtract 15 from var, then we should copy 15 to temp variable and then performing Reaction 5.1.

\subsection{Sample Compound Reaction - Multiplication}
My compiler directly supports $*$ operator for multiplication. However, to implement multiplication, we followed this routine for multiplication of two numbers x and y. 

\begin{lstlisting}
while (x > 0):
	result = result + y;
	x = x - 1;
\end{lstlisting}

\begin{figure}[ht!]
\centering
\includegraphics[scale=0.3]{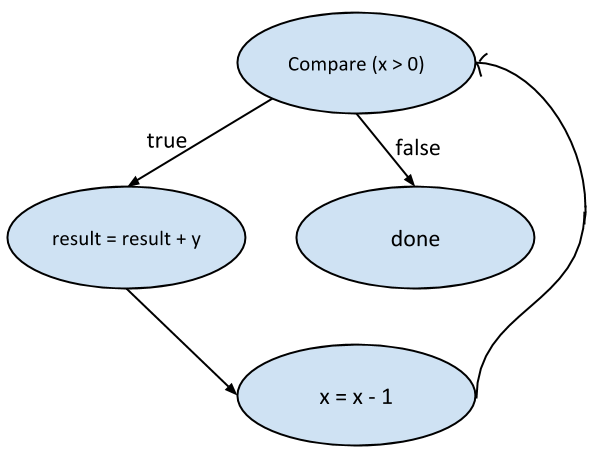}
\centering
\caption{The work-flow of multiply operator inside compiler.}
\label{imageMulWorkflow}
\end{figure}

To implement this, we need three sets of reactions: Comparison, Copy, and Decrement. The comparison reactions should provide a start signal to copy reactions and decrement reactions, if its condition is met. The flow is shown in Fig.~\ref{imageMulWorkflow}.

Please refer supplementary files to see complete set of reactions for multiply module.

\section{Compiler Discussion - Parsing and Operation handling}
In this compiler, we have adopted python like syntax along with parenthesis instead of indentation. As of now, the compiler supports following operations: 

\begin{itemize}
\item Declaration and assignment of a variable whose name is of the form $[a-z][a-z0-9]^+$.
\item Addition operation.
\item Subtraction operation.
\item Multiplication operation.   
\end{itemize}

The compiler also supports following operations, however, it is not yet able to generate the xml code for these operations. This means that compiler recognizes the syntax for these operations, generates an Abstract Syntax Tree (AST) for them, and is also able to travserse it for these operations. 

\begin{itemize}
\item Division operator.
\item All the types of comparison operations: $<$, $>$, $<=$, $>=$, $==$, and $!=$.
\item The "if" branching statement
\item The "while" looping statement.
\end{itemize}

Some of the important points to be noted about compiler are as under: 

\begin{itemize}
\item Since we are following stochastic model for CRN, the coefficients for any reactant or product will always be non-negative integer. So, our compiler only has integer variables as of now. One needs to come up with some kind of encoding to represent negative numbers and decimal numbers using CRN.

\item My compiler replicates Intel's IA32 architecture in terms of register. It consists of 3 register like variables $\_ localx$, $\_ localy$, and $\_ localz$ just like IA32 consists of register eax, ebc, ecx and edx.

\item At any given point of time, we store temporary value of result, which can be used later, in $\_ localz$. The other registers are used to store temporary value inside a given operation i.e intra-operation. Say for example, while performing multiplication 2 * 3, we store 2 and 3 in $\_localx$ and $\_localy$ respectively while their result is stored in $\_ localz$.

\item To assign a value to variable, we need to quantify the value of a constant. Say for example, to perform $x = 10$, we define a specie called one\_zero whose amount is 10 and then perform copy reaction on x. 

\item We have used a hash map data structure for symbol table in our compiler. As of now, there is nothing like scope of variable i.e every variable is global but this can easily be implement by using a nested hash map inside a hash map.

\item The compiler works in two different modes: Interactive mode as well as File mode. In interactive mode, the compiler give line wise output as if it is an interpreter while in file mode, the compiler takes a file as input and gives single output.

\item The way in which Flex and Bison interact with GCC to create executable for the compiler is shown in Fig.~\ref{imageWorkflow}.

\item The associativity and precedence of the operators available in compiler is shown in Table.1.
\end{itemize}

\begin{table}
\small
    \label{tableOpr}
    \centering
    \caption{}
    \begin{tabular}{ |l| l|l |l|}
    \hline
    Precedence & Operator & Associativity \\ \hline \hline
    1 & *, / & Left \\ \hline
    2 & +, - & Left \\ \hline
    3 & $<$, $>$, $<=$, $>=$, $=$, $!=$ & Left \\    
    \hline
    \end{tabular}   
\end{table}

\section{Intermediate Output}
\begin{figure}[ht!]
\includegraphics[width=0.4\textwidth]{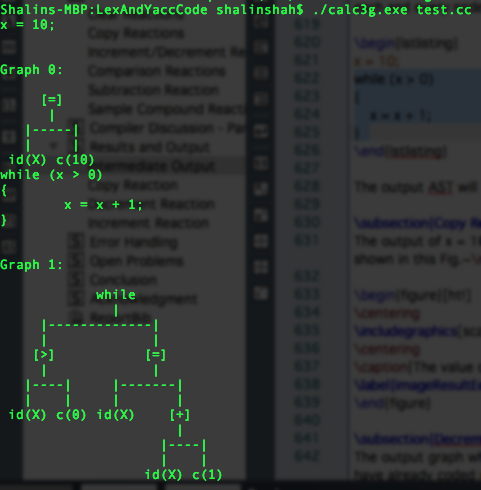}
\centering
\caption{AST for x =10 and while loop.}
\label{imageResultInter}
\end{figure}

After the syntax checking stage, the compiler generates an AST for the code and it returns the head pointer of this AST. Then, by traversing each and every node of this AST, we can generate the chemical reactions for the same. For example, consider the following code: 

\begin{lstlisting}
x = 10;
while (x > 0)
{
	x = x + 1;
}	
\end{lstlisting}

The output AST will be as shown in Fig.~\ref{imageResultInter}.

\section{Final Output}
The XML output is too large to be shown here. However, the simulated graph output is shown. 

\subsection{Copy Reaction}
\begin{figure}[ht!]
\centering
\includegraphics[scale=0.25]{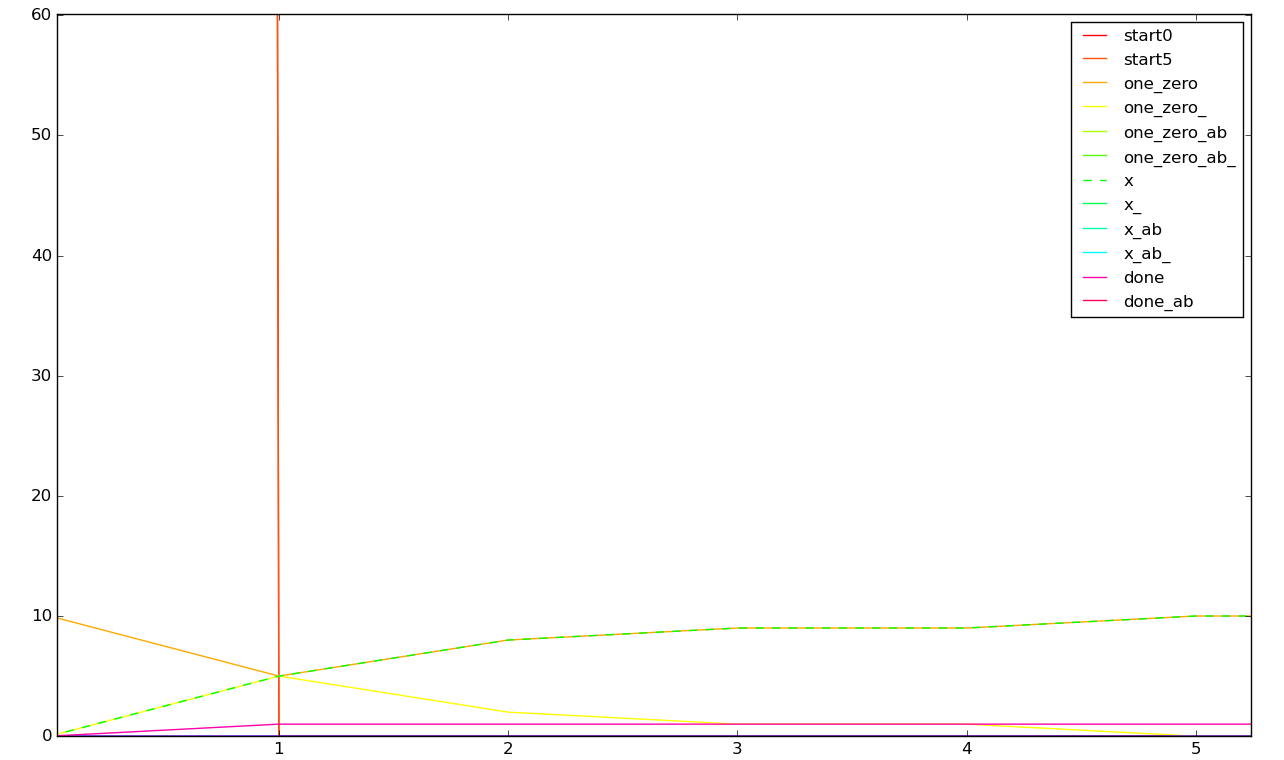}
\centering
\caption{The value of variable x increases with time and after it reaches 10, it becomes constant over time.}
\label{imageResultEq}
\end{figure}

The output of x = 10 will generate 10 reactions and 18 species after compilation. This file when simulated in CAIN will generate output graph as shown in this Fig.~\ref{imageResultEq}.

\subsection{Decrement Reaction}
\begin{figure}[ht!]
\centering
\includegraphics[scale=0.25]{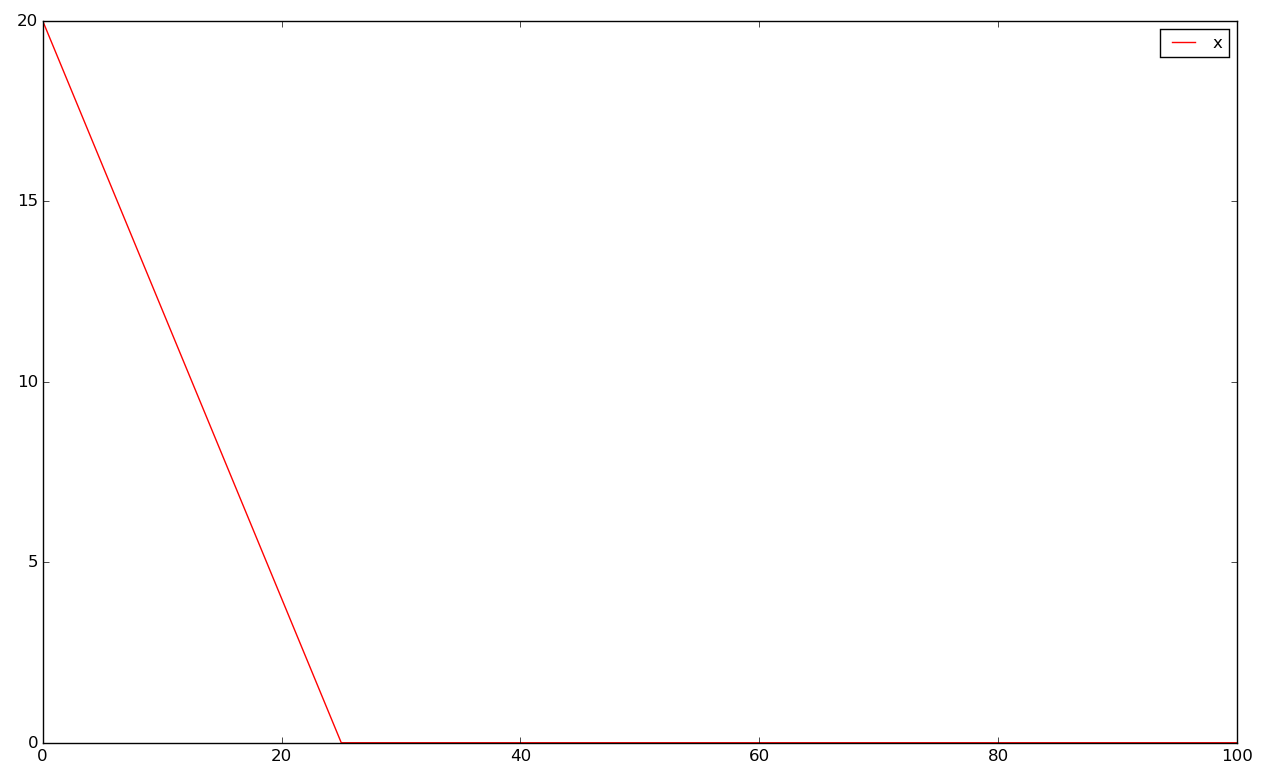}
\centering
\caption{The value of variable x decreases with time and after it reaches 0, it becomes constant over time.}
\label{imageResultDec}
\end{figure}

The output graph for decrement reaction on x = 20 when simulated in CAIN will generate output graph as shown in Fig.~\ref{imageResultDec}. Note that compiler doesn't support unary subtract or add symbol yet. 

\subsection{Increment Reaction}
The output graph of increment reaction will be similar to decrement with the only difference being sign of slope. This graph will have positive slope unlike decrement graph. 

\section{Error Handling}
Like any compiler, our compiler also issues very basic errors. As of now, the compiler is able to catch, mainly, syntax errors. Say for example, $x = 20.22$ will generate a syntax error since the compiler is only supporting integer variables or forgetting a ';' at the end of a statement will generate a syntax error. Also, the compiler return line number for the which error was reported. For example, consider this erroneous code: 

\begin{lstlisting}
x = 10;
y = 20
z = x + y;
\end{lstlisting}

When we execute this code, compiler reports - \textit{syntax error at/near line 2}. Other than syntax errors, compiler also reports undeclared variable error and insufficient memory error in case it is out of memory. 

\section{Open Problems}
While trying to develop compiler, we were able to identify following open problems (currently, we are working on them): 

\begin{itemize}
\item An encoding method to represent negative numbers. we came up with an idea of using 9's complement method but the problem with this method is highest decimal has to be discarded once it overflows. How to do that using CRN is still a question who's answer we are seeking.

\item If one is able to come up with encoding for negative numbers, next step should be to find encoding for real numbers. 

\item Inter-compatibility of reactions. Suppose, we do following operations: 

\begin{lstlisting}
x = 10;
y = x;
\end{lstlisting}

In this case,  value of x will be copied to y after copy reactions for x. This creates weird behavior since we need to address the problem of synchronizing \textit{rate independent} reactions. The eerie behavior is due to start and done signal of reactions and their quantities. A method to do this has been proposed in \cite{seenumReport} but it doesn't address the issue completely. Deciding the optimal initial quantity of absence indicator, start and done signals still remains open. 
\end{itemize}

\section*{Supplementary File}
The supplementary software for this work can be found on this Github link:

\url{https://github.com/shalinshah1993/CCompilerX}

It contains sample codes for addition, multiplication and subtraction generated by the compiler.

\section*{Conclusion}
With things such as very low probability of error, efficient Turing Universal behavior, and energy efficiency reactions, CRN is surely a great alternative to silicon processors. Therefore, a need of compiler is imperative. However, there are still several challenges, a few of which are mentioned in a section above, which one needs to address before this dream of compiler can be realized. That being said, this compiler is at such a stage that a strong base is developed on which one can further this work simply by adding modules.

\section*{Acknowledgment}
We would like to thank Krishna Gopal for healthy discussions and suggestions. We would also like to thank Frank Thomas Braun for his tree generation library. 

\ifCLASSOPTIONcaptionsoff
  \newpage
\fi

\bibliographystyle{IEEEtran}
\bibliography{ReportBib}

\end{document}